\documentclass[5p]{elsarticle}

\usepackage{lineno,color,bm}
\usepackage{tabularx, dcolumn}
\usepackage{amsmath,arydshln,amssymb,wasysym}
\newcolumntype{d}{D{+}{\,\pm\,}{6,3}}
\modulolinenumbers[5]

\usepackage[subrefformat=parens]{subcaption}
\journal{Journal of \LaTeX\ Templates}

\begin{document}

\begin{frontmatter}

\title{Development of co-located $^{129}$Xe and $^{131}$Xe nuclear spin masers with external feedback scheme}

\author[1,2]{T.~Sato\corref{cor1}}
\ead{tomoya.sato@riken.jp}
\author[1,2]{Y.~Ichikawa}
\author[1]{S.~Kojima}
\author[1]{C.~Funayama}
\author[1]{S.~Tanaka}
\author[3,4]{T.~Inoue}
\author[4]{A.~Uchiyama}
\author[2,5]{A.~Gladkov}
\author[2]{A.~Takamine}
\author[1]{Y.~Sakamoto}
\author[1]{Y.~Ohtomo}
\author[1]{C.~Hirao}
\author[1]{M.~Chikamori}
\author[1]{E.~Hikota}
\author[1]{T.~Suzuki}
\author[1]{M.~Tsuchiya}
\author[6]{T.~Furukawa}
\author[7]{A.~Yoshimi}
\author[8]{C.P.~Bidinosti}
\author[9]{T.~Ino}
\author[2]{H.~Ueno}
\author[10]{Y.~Matsuo}
\author[11]{T.~Fukuyama}
\author[12]{N.~Yoshinaga}
\author[13,4]{Y.~Sakemi}
\author[1,2]{K.~Asahi}

\address[1]{Department of Physics, Tokyo Institute of Technology, 2-12-1 Oh-okayama, Meguro, Tokyo 152-8551, Japan}
\address[2]{RIKEN Nishina Center for Accelerator-Based Science, 2-1 Hirosawa, Wako, Saitama 351-0198, Japan}
\address[3]{Frontier Research Institute for Interdisciplinary Sciences (FRIS), Tohoku University, 6-3 Aoba, Aramaki, Aoba, Sendai 980-8578, Japan}
\address[4]{Cyclotron and Radioisotope Center (CYRIC), Tohoku University, 6-3 Aoba, Aramaki, Aoba, Sendai 980-8578, Japan}
\address[5]{Department of Physics, Kyungpook National University, 80 Daehakro, Bukgu, Daegu 41566, Korea}
\address[6]{Department of Physics, Tokyo Metropolitan University, 1-1 Minami-Ohsawa, Hachioji, Tokyo 192-0397, Japan}
\address[7]{Research Institute for Interdisciplinary Science (RIIS), Okayama University, 3-1-1 Tsushimanaka, Kita, Okayama 700-8530, Japan}
\address[8]{Department of Physics, University of Winnipeg, 515 Portage Avenue, Winnipeg, MB, R3B 2E9, Canada}
\address[9]{Institute of Materials Structure Science (IMSS), High Energy Accelerator Research Organization (KEK), Tsukuba, Ibaraki 305-0801, Japan}
\address[10]{Department of Advanced Sciences, Hosei University, 3-7-2 Kajino-cho, Koganei, Tokyo 184-8584, Japan}
\address[11]{Research Center for Nuclear Physics (RCNP), Osaka University, 10-1 Mihogaoka, Ibaraki, Osaka 567-0047, Japan}
\address[12]{Graduate School of Science and Engineering, Saitama Unicersity, 255 Shimo-Okubo, Sakura-ku, Saitama City, Saitama 338-8570, Japan}
\address[13]{Center for Nuclear Study (CNS), Graduate School of Science, University of Tokyo, 7-3-1 Hongo, Tokyo 113-0033, Japan}

\cortext[cor1]{Corresponding author}





\begin{abstract}
We report on the operation of co-located $^{129}$Xe and $^{131}$Xe nuclear spin masers
with an external feedback scheme,
and discuss the use of $^{131}$Xe as a comagnetometer in measurements of the $^{129}$Xe spin precession frequency.
By applying a correction based on the observed change in the $^{131}$Xe frequency,
the frequency instability due to magnetic field and cell temperature drifts are eliminated by two orders of magnitude.
The frequency precision of 6.2~$\mu$Hz is obtained for a $10^4$~s averaging time,
suggesting the possibility of future improvement to $\approx$ 1~nHz by improving the signal-to-noise ratio of the observation.
\end{abstract}


\end{frontmatter}

\section{Introduction}
\label{sec:intro}
The ability to measure the precession frequency of a nuclear spin figures prominently
in many experiments in low-energy frontier of fundamental physics,
such as the search for axion-like particles~\cite{Tullney2013,Bulatowicz2013,Vasilakis2009},
violations of the Lorentz invariance~\cite{Bear2000,Allmendinger2014} and the time-reversal symmetry~\cite{Rosenberry2001,Graner2016,Pendlebury2015}.
Noble-gas atoms are especially appropriate for such measurements because of their desirable properties:
long spin coherence times by virtue of the closed-shell electronic configuration
and high degrees of spin polarization achievable by means of the spin-exchange optical pumping (SEOP) technique~\cite{Walker1997}.

To achieve precision measurements of spin precession frequency,
we have developed a nuclear spin maser
with an optical spin detection and an artificial feedback framework~\cite{Yoshimi2002,Yoshimi2012,Ichikawa2014,Sato2015,Inoue2016}.
The spin maser enables the elongation of the spin precession for long time durations,
far beyond the transverse relaxation time $T_2$.
The spin maser is one of the possible schemes
in experiments which require long-term observation of the spin precession
such as searches for the electric dipole moment (EDM) and the oscillating EDM induced by axion-matter coupling
in a low-frequency region~\cite{Flambaum2016,Budker2014,Abel2017}.
In addition, the artificial feedback scheme allows the operation of maser at very low field
and also the freedom to tune the gain factor for maser.
This feature is advantageous for stabilization of the maser operation conditions as discussed later.

In previous work~\cite{Inoue2016},
we achieved a one-shot frequency precision of 7.9~nHz for an averaging time of $3\times10^4$~s
by using a $^{129}$Xe spin maser,
but at the same time we found that
the maser frequency in repeated measurements varied in a range of mHz order.
This frequency instability turned out to occur from the frequency drifts
associated with long-term changes in the operational conditions.
In order to eliminate changes of the precession frequency arising from long-term drifts in the magnetic field
which often is the major cause of the systematic uncertainty,
a ``comagnetometer'' is widely used.
The change in the precession frequency of the spin of interest is canceled out
by the frequency comparison with another spin (comagnetometer) occupying the same location,
provided that the two spins sense the common environment.
For a comagnetometer for $^{129}$Xe, the $^3$He spin is often used~\cite{Tullney2013,Bear2000,Allmendinger2014,Rosenberry2001,Ichikawa2014,Sato2015}
because it would be quite appropriate as a reference due to its long spin coherence times
(\textit{e.g.} $T_2 > $100~hours is reported~\cite{Heil2013}).
However, the problem of a frequency shift caused by the Fermi contact interaction
between a noble gas atom and a polarized Rb atom~\cite{Schaefer1989}
(which is an essential ingredient of the SEOP technique) remains to be solved
because the magnitude of the interaction largely differs for $^{129}$Xe and $^3$He~\cite{Ma2011,Romalis1998}.

To eliminate systematic uncertainty arising from the contact interaction,
we employ $^{131}$Xe instead of $^3$He as a comagnetometer for the $^{129}$Xe experiment.
Since the magnitude of the contact interaction between $^{131}$Xe and Rb is very similar to that between $^{129}$Xe and Rb,
the effect of contact interaction can be greatly reduced.
This advantage has been demonstrated in the search for a new scalar-pseudoscalar interaction
conducted by comparing free-induction-decay frequencies of the two Xe isotopes~\cite{Bulatowicz2013}.
The shortened spin coherence time of $^{131}$Xe due to quadrupole relaxation~\cite{Kwon1981,Stupic2011} is one of the issues
that need consideration in the measurement using $^{131}$Xe,
because it may limit the improvement of the statistical sensitivity.
This difficulty will be overcome by introducing a maser scheme for $^{131}$Xe spin, as well.

In this paper, we report the first operation of
co-located $^{129}$Xe and $^{131}$Xe nuclear spin masers
with an external feedback scheme.
Brief introduction of the nuclear spin maser with external feedback,
the setup, and developments for improved stabilities of the operational conditions
and experimental procedure are given in Section 2.
The frequency characteristics and long-term stabilities of the masers are reported and discussions are presented in Section 3.
The conclusion is given in Section 4.

\section{\texorpdfstring{Operation of co-located $^{129}$Xe/$^{131}$Xe masers}{Operation of co-located 129Xe/131Xe masers}}
\label{sec:operation}
The motion of spins in a magnetic field ${\bm B} = (B_x,B_y,B_z)$ is described by modified Bloch equations
which include effect of a feedback field to maintain the maser oscillation:
\begin{align}
    \frac{{\rm d}P_x(t)}{{\rm d}t} &= \gamma [ P_y(t) B_z - P_z(t) B_y(t) ] - \frac{P_x(t)}{T_2}, \label{eq:Bloch_x} \\
    \frac{{\rm d}P_y(t)}{{\rm d}t} &= \gamma [ P_z(t) B_x(t) - P_x(t) B_z ] - \frac{P_y(t)}{T_2}, \label{eq:Bloch_y} \\
    \frac{{\rm d}P_z(t)}{{\rm d}t} &= \gamma [ P_x(t) B_y(t) - P_y(t) B_x(t) ] - \frac{P_z(t) - P_0}{T_1^*}. \label{eq:Bloch_z}
\end{align}
Here, ${\gamma}$ is the gyromagnetic ratio of the spin species;
$P_i(t)$ with $i=x,y,z$ is the $i$-th component of polarization;
$T_1^*$ is the effective longitudinal relaxation time
of the maser species including the pumping effect due to the SEOP;
$T_2$ is the transverse relaxation time;
$B_i(t)$ is the $i$-th component of the magnetic field;
$P_0$ is the constant value of polarization that is attained in the $z$-direction
when the equilibrium is reached.
We assume that the $z$-component $B_z$ of ${\bm B}$ is static and the largest.
The basic principles of operation of the active feedback spin maser are described in the previous papers~\cite{Yoshimi2002,Yoshimi2012,Inoue2016}.
The spin precession of the Xe atom is maintained by the application of the feedback field
which is generated according to the precession signal of the Xe itself.
In the current experiment,
we introduce a novel ``fixed-amplitude'' feedback scheme in order to stabilize the maser behavior
against changes in the conditions of the optical spin detection.

To see how this works, let us first consider the normal operation of the active spin maser.
In addition to a static field ${\bm B}_0 = (0,0,B_0)$,
we apply a transverse field ${\bm B}_{\rm FB}(t) = (B_x^{\rm FB}(t),B_y^{\rm FB}(t),0)$
whose azimuthal angle is $- 90^{\circ}$ shifted
from that of the transverse polarization ${\bm P}_{\rm T}(t) = (P_x(t),P_y(t),0)$:
\begin{align}
    B_x^{\rm FB}(t) &= \frac{1}{\gamma \tau_{\rm FB}} \frac{P_y(t)}{P_0}, \label{eq:B_x_normal} \\
    B_y^{\rm FB}(t) &= - \frac{1}{\gamma \tau_{\rm FB}} \frac{P_x(t)}{P_0} \label{eq:B_y_normal},
\end{align}
where $\tau_{\rm FB}$ is a constant characterizing the strength of the feedback field in the active spin maser.
By setting $P_x + iP_y = P_{\rm T} e^{-i \gamma B_0 t}$,
the set of equations (\ref{eq:Bloch_x})-(\ref{eq:Bloch_z}) reduces to
\begin{align}
    \frac{d P_{\rm T}(t)}{dt} &= \left( \frac{1}{\tau_{\rm FB}} \frac{P_z(t)}{P_0} - \frac{1}{T_2} \right) P_{\rm T}(t), \label{eq:pol_T_active} \\
    \frac{d P_z(t)}{dt} &= - \frac{1}{\tau_{\rm FB}} \frac{P_{\rm T}(t)^2}{P_0} + \frac{P_0 - P_z(t)}{T_1^*}. \label{eq:pol_z_active}
\end{align}
${{\bm P}_{\rm T}(t)}$ actually is the magnitude of the transverse polarization
which rotates on the $xy$-plane at an angular frequency $- \gamma B_0$.
The $P_{\rm T}$ and $P_z$ in the stationary state at $t \rightarrow \infty$ are
\begin{align}
    P_{\rm T}^{\rm eq} &= \tau_{\rm FB} \sqrt{\frac{1}{T_1^*} ( \frac{1}{\tau_{\rm FB}} - \frac{1}{T_2} )} P_0,\\
    P_z^{\rm eq} &= \frac{\tau_{\rm FB}}{T_2} P_0.
\end{align}
The strength of the feedback field in this situation is given by
\begin{align}
    B_{\rm FB}^{\rm eq} &= \frac{1}{\gamma \tau_{\rm FB}} \frac{P_{\rm T}^{\rm eq}}{P_0} 
    = \frac{1}{\gamma} \sqrt{\frac{1}{T_1^*} \left( \frac{1}{\tau_{\rm FB}} - \frac{1}{T_2} \right)}.
\end{align}
When $\tau_{\rm FB}$ is set to be
\begin{equation}
    \tau_{\rm FB}^{\rm max} = \frac{T_2}{2},
    \label{eq:T2_max_active}
\end{equation}
$P_{\rm T}^{\rm eq}$ is maximum;
\begin{align}
    P_{\rm T}^{\rm eq} &= \frac{1}{2} \sqrt{\frac{T_2}{T_1^*}} P_0, \label{eq:pol_T_eq_active} \\
    P_z^{\rm eq} &= \frac{1}{2} P_0. \label{eq:pol_z_eq_active}
\end{align}
The feedback field corresponding to $\tau_{\rm FB}^{\rm max}$ is
\begin{equation}
    B_{\rm FB}^{\rm max} = \frac{1}{\gamma} \frac{1}{\sqrt{T_1^* T_2}}.
\end{equation}

Although the amplitude of ${\bm B}_{\rm FB}(t)$ intended
in Eqs.(\ref{eq:B_x_normal}) and (\ref{eq:B_y_normal}) is proportional to the magnitude of ${\bm P}_{\rm T}(t)$,
the spin maser can also be operated by keeping the amplitude for ${\bm B}_{\rm FB}(t)$ constant in time.
In this case, the frequency instability originating from the change in the amplitude for ${\bm B}_{\rm FB}(t)$
can be eliminated.
Thus in the active spin maser in an amplitude-fixed mode,
${\bm B}_{\rm FB}(t)$ rotates in synchronization with ${\bm P}_{\rm T}(t)$
but its magnitude is kept constant intentionally.
The amplitude-fixed feedback fields $B_x^{\rm fixed}(t)$ and $B_y^{\rm fixed}(t)$ are written respectively as
\begin{align}
    B_x^{\rm fixed}(t) &= \frac{1}{\gamma \tau_{\rm FB}} \frac{P_y(t)}{\sqrt{P_x(t)^2+P_y(t)^2}}, \label{eq:B_x_fixed} \\
    B_y^{\rm fixed}(t) &= - \frac{1}{\gamma \tau_{\rm FB}} \frac{P_x(t)}{\sqrt{P_x(t)^2+P_y(t)^2}}. \label{eq:B_y_fixed}
\end{align}
Thus, the equations of motion in the rotating frame for the maser with the amplitude-fixed feedback field
in the rotating frame are given as
\begin{align}
    \frac{d P_{\rm T}(t)}{dt} &= \left( \frac{1}{\tau_{\rm FB}} \frac{P_z(t)}{P_{\rm T}(t)} - \frac{1}{T_2} \right) P_{\rm T}(t), \label{eq:pol_T_fixed} \\
    \frac{d P_z (t)}{dt} &= - \frac{1}{\tau_{\rm FB}} P_{\rm T}(t) + \frac{P_0 - P_z(t)}{T_1^*}. \label{eq:pol_z_fixed}
\end{align}
$P_{\rm T}$ and $P_z$ in equilibrium are
\begin{align}
    P_{\rm T}^{\rm eq} &= \frac{T_2 \tau_{\rm FB}}{\tau_{\rm FB}^2 + T_1^* T_2} P_0,\\
    P_z^{\rm eq} &= \frac{\tau_{\rm FB}^2}{\tau_{\rm FB}^2 + T_1^* T_2} P_0.
\end{align}
The value of $\tau_{\rm FB}$ which maximizes $P_{\rm T}$,
and the corresponding values of $P_{\rm T}^{\rm eq}$ and $P_z^{\rm eq}$ are
\begin{align}
    \tau_{\rm FB}^{\rm max} &= \sqrt{T_1^* T_2}, \label{eq:T2_max_fixed} \\
    P_{\rm T}^{\rm eq} &= \frac{1}{2} \sqrt{\frac{T_2}{T_1^*}} P_0, \label{eq:pol_T_eq_fixed} \\
    P_z^{\rm eq} &= \frac{1}{2} P_0. \label{eq:pol_z_eq_fixed}
\end{align}
The strength of the feedback field corresponding to the value of $\tau_{\rm FB}$ above is expressed as
\begin{align}
    B_{\rm FB}^{\rm eq} = \frac{1}{\gamma \tau_{\rm FB}^{\rm max}} 
     = \frac{1}{\gamma \sqrt{T_1^* T_2}}.
\end{align}

\begin{figure}[htb]
  \centering
    \includegraphics[width=8.0cm]{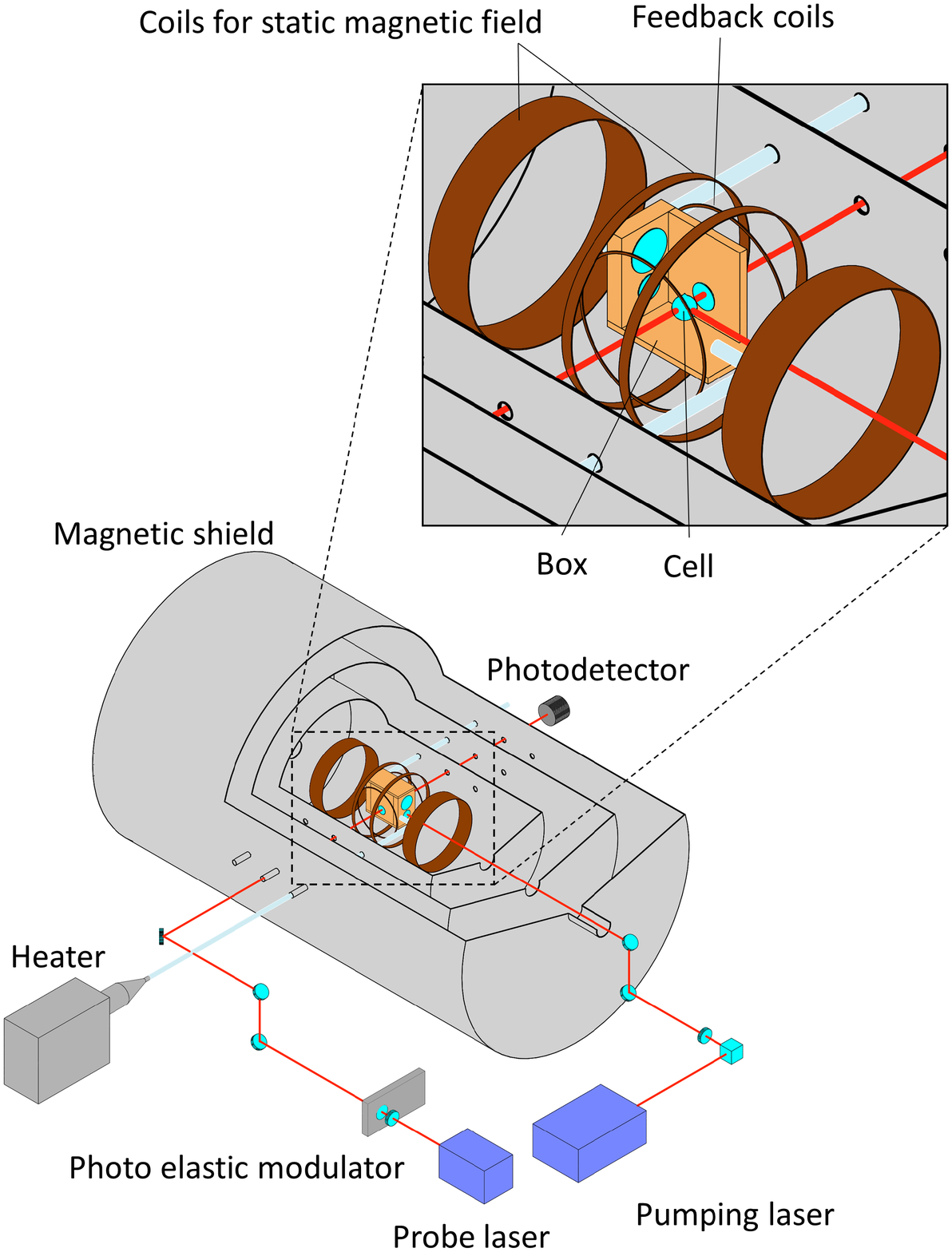}
    \caption{A schematic view of the experimental setup for co-located $^{129}$Xe and $^{131}$Xe nuclear spin masers.
                A gas cell filled with $^{129}$Xe, $^{131}$Xe, $^3$He, N$_2$ gases, and Rb vapor, was set in a box in which temperature was stabilized.
                The box was placed in a set of coils which generated a static magnetic field $B_0$, and was enclosed in a three-layered magnetic shield.
                A circularly polarized light for SEOP was introduced to the cell parallel to $B_0$.
                A transmission of a probe light whose direction was orthogonal to $B_0$ was monitored by a photodetector.}
    \label{fig:setup}
\end{figure}

The schematic view of the setup is shown in Fig.~\ref{fig:setup}.
In order to study the basic properties of the $^{131}$Xe maser,
a spherical cell made of Pyrex glass is chosen so that the quadrupole splitting would disappear.
The outer diameter of the cell is 20~mm and the wall thickness is 1~mm.
Before the gas filling process, the cell was cleaned using a neutral detergent (FineClean70), acetone, ethanol and nitric acid.
Thus cleaned cell was baked at 200~$^\circ$C for a few days in a vacuum of a level of $10^{-6}$~Torr.
The cell contained 1~Torr of $^{129}$Xe, 25~Torr of $^{131}$Xe, 10~Torr of N$_2$, 425~Torr of $^3$He and a drop of Rb metal.
The gases except for N$_2$ were isotopically enriched ($>$99.9~\%) ones.
Their partial pressures were chosen so as to maximize the magnetization of $^{131}$Xe.

\begin{figure}[htb]
  \centering
    \includegraphics[width=8.0cm]{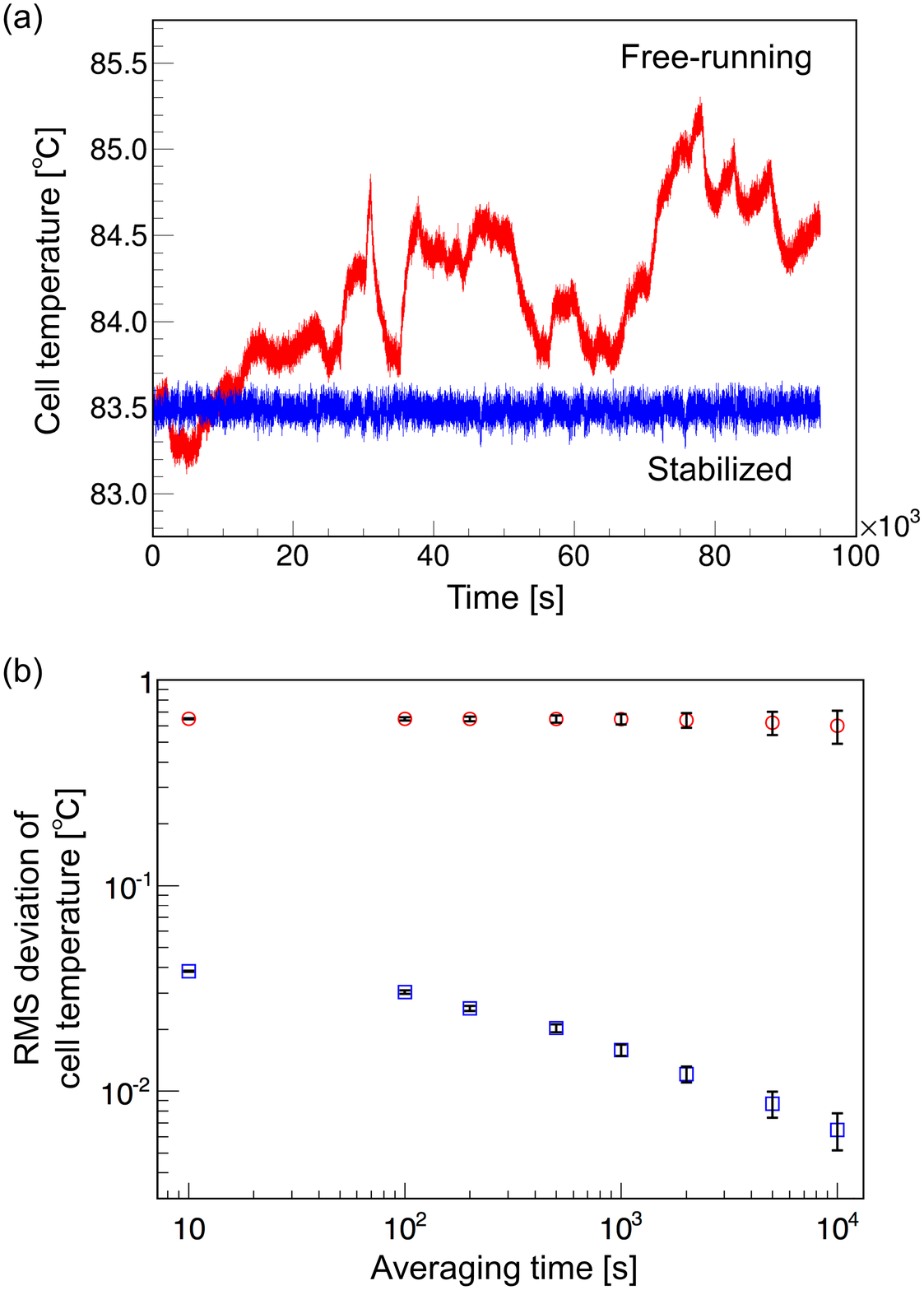}
    \caption{(a) Time series plot of the cell temperature.
                Red and blue lines show the temperature before and after the stabilization.
                (b) Plot of the rms deviation of the cell temperature as a function of averaging time.
                Red open circles and blue open squares show the data before and after the stabilization, respectively.}
    \label{fig:temp_stab}
\end{figure}

The cell was set in a box made of polyetheretherketone (PEEK).
The box had an AR-coated window on each side for the laser access
and had air intake/outgo ports for a heated air flow.
The temperature around the cell was maintained at $\sim$85$^{\circ}$C by a PID-controlled heater (Hotwind System, Leister Technologies).
The control loop for the temperature stabilization consisted of the heater, a temperature indicator/PID-controller (LT370, Chino Corp.)
and a platinum resistance temperature detector (RTD).
The stability of the temperature around the cell was evaluated.
The temperature was measured by the RTD attached to the bottom of the cell.
Figure~\ref{fig:temp_stab} shows the time series plot of the cell temperature
and its root-mean-square (rms) deviation as a function of averaging time.
For an averaging time of $10^4$~s,
the rms deviation of the cell temperature was found to be 0.0065$\pm$0.0013$^{\circ}$C.
Two sets of Helmholtz coils, which provided the feedback field for the cell, surrounded the box.
The static magnetic field and the axis of the Helmholtz coils were orthogonal to each other.

The box and coils for the feedback field were set in a magnetic shield
so that they were decoupled from the residual magnetic field,
in order to reduce the drift and fluctuation of the residual magnetic field.
The magnetic shield consisted of three nested Permalloy cylinders 2-mm thick each.
Their sizes were 800-mm$\phi \times$  1300 mm, 600-mm$\phi \times$ 1000 mm and 400-mm$\phi \times$ 680 mm in order from outermost to innermost.
They were individually capped with end caps, each of which had a 50-mm hole on the symmetry axis for the introduction of a pumping laser light.
Each layer had three 20-mm holes on both left and right sides for the probe light, cables, and air tubes access.
The shielding factor of the shield was estimated at a level of $10^4$.
An active field cancellation system~\cite{Inoue2016} surrounded the magnetic shield
in order to cancel the residual field along the symmetry axis.
The system consisted of three rectangular coils of sizes 360~cm (horizontal) $\times$ 225~cm (vertical) each.
The three coils were spaced with intervals of 145~cm.
The current fed to the cancellation coils was controlled by a LabVIEW based controller
so that the magnetic field (measured with a fluxgate magnetometer)
outside the shield at a distance of 50~cm from the wall of the outermost shield cylinder was maintained stable.

Inside the innermost layer of the shield, a set of coils to provide $B_0$ was installed.
The $B_0$ coils were designed so that the second and the fourth order components of the field
were kept very small near the central area~\cite{Sakamoto2015}.
In the current experimental condition, the masers were operated typically at a static magnetic field of $B_0 \sim$ 9.38~mG,
at which the precession frequency was calculated to be $\sim$11~Hz for $^{129}$Xe and $\sim$3~Hz for $^{131}$Xe.
The electrical current for the $B_0$ coil was supplied by a current source stabilized by a feedback loop~\cite{Inoue2016}.

For the SEOP, a DFB laser combined with a tapered amplifier (TA) provided 795 nm laser light (pumping light)
parallel to the $B_0$ field to optically pump Rb in the cell.
The output power of the laser system (TOPTICA, TA-DFB) was typically 1~W.
The laser light was circularly polarized by a $\lambda/4$ wave plate.
Another DFB laser provided a light for the optical detection of the spin precession (probe light)
perpendicular to $B_0$.
The typical power for the maser operation was from a few to 10~mW.
The transmission of the light through the cell was monitored by a photodetector (PD).
A photo-elastic modulator (PEM) was inserted in the light path which modulated the light helicity at 50~kHz
and sent a reference signal in synchronization with the helicity modulation to a lock-in amplifier which we refer to later.

\begin{figure}[htb]
  \centering
    \includegraphics[width=7.4cm]{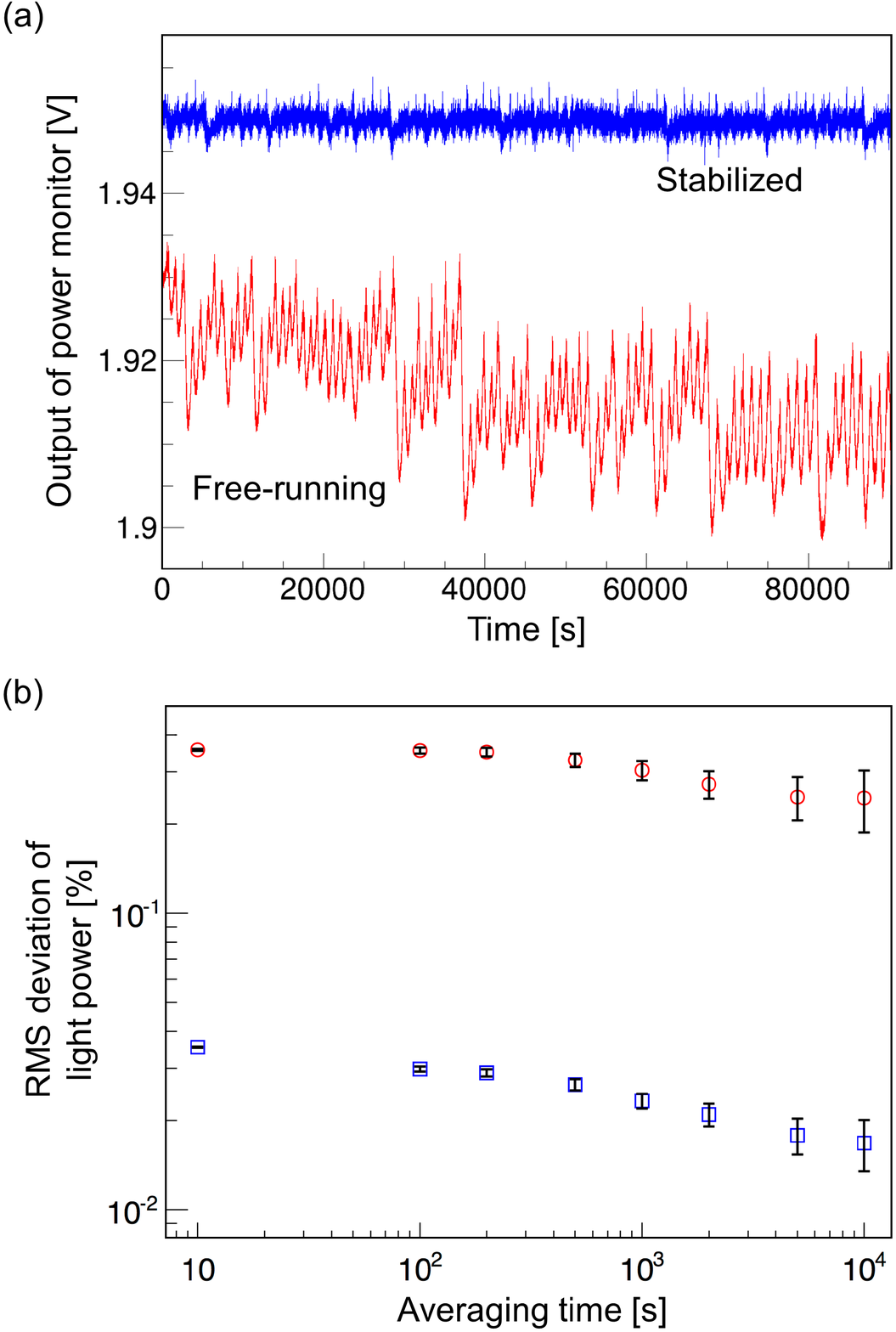}
    \caption{(a) Time series plot of the power monitor.
                Red and blue lines show the outputs before and after the stabilization.
                (b) Plot of the rms deviation of the output of the power monitor as a function of the averaging time.
                Red open circles and blue open squares show the free-running and stabilized data, respectively.}
    \label{fig:power_stab}
\end{figure}

In order to eliminate long-term drifts of the light power which would cause changes of the polarization and number density of Rb atoms,
we incorporated a feedback control for the power stabilization of the lasers.
The control system consisted of a Glan-Laser polarizer (GLP),
a rotatable $\lambda/2$ plate, a beam sampler and a PD.
The polarization plane of the laser light was controlled by the rotatable wave plate.
According to the rotation angle with respect to the fast axis of the GLP,
the output power from the GLP located just after the wave plate was changed.
A part of the laser light was sampled by using the beam sampler.
The rotation angle of the wave plate was controlled by the LabVIEW-based feedback program
so that the power of the sampled light became stable.
Figure~\ref{fig:power_stab} shows a time series plot of the output of the light power monitor
and its rms deviation plotted as a function of averaging time.
For an averaging time of $10^4$~s,
the rms deviation was approximately 1.7$\times 10^{-2}$~$\%$.

To generate the fixed-amplitude feedback field,
the analog feedback module that had been used in Ref.~\cite{Inoue2016} was replaced by a digital one.
In the digital feedback module, the feedback signal was processed by a CPU (Arduino Uno).
With C-language based programing capability,
the new feedback system provided flexible signal processing.
In addition,
the use of the digitalized feedback instead of the analog one reduced the jitter and drift in the phase of the feedback signal,
thus eliminating the maser instability arising from the ``pulling effect''~\cite{Yoshimi2002}.

The procedure to operate the active feedback nuclear spin maser is as follows:
Under the static field $B_0$, Xe spin is polarized via SEOP with Rb.
By a thermal fluctuation or by application of a pulsed magnetic field,
the polarized Xe nuclear spin starts precession.
The transverse component of the Xe spin is transferred to the Rb atomic spin,
and thus the Xe precession signal is detected as a sinusoidal change of the probe light transmission~\cite{Green1998}. 
The signal was lock-in amplified in two stages for the improvement of the signal-to-noise ratio (SNR).
The signal from the photodetector which represents the probe light power is sent
to the first lock-in amplifier (LA1) where its frequency component at the frequency of the PEM modulation is amplified.
The output signal of LA1 is sent to the second lock-in amplifier (LA2)
where its component at the Xe precession frequency is selected.
The outputs from LA2 are sent to a circuit for the feedback field generation and also to an analog-to-digital converter.
The feedback field $B_{\rm FB}$ needed to maintain the spin precession is generated in the following way:
The outputs of LA2 represent the beat signals between the precession signal
and a signal from a function generator whose frequency is detuned by a few to a few hundred mHz from the calculated precession frequency.
Thus the beat signal $V_{\rm X}(t)$ between the precession signal $V_{\rm sig}(t) = V_0 \sin{(2 \pi \nu_0 t + \phi_0)}$
and the reference signal $V_{\rm ref1}(t) = V_{\rm r} \sin{(2 \pi \nu_{\rm r} t + \phi_{\rm r})}$ is written as
\begin{equation}
    V_{\rm X}(t) = \frac{1}{2}V_0 V_{\rm r} \cos{[2 \pi (\nu_0 - \nu_{\rm r})t + (\phi_0 + \phi_{\rm r})]}.
\end{equation}
LA2 furnishes another output $V_{\rm Y}(t)$ whose phase is shifted by $+$90$^\circ$ from $V_{\rm X}(t)$;
\begin{equation}
    V_{\rm Y}(t) = \frac{1}{2}V_0 V_{\rm r} \sin{[2 \pi (\nu_0 - \nu_{\rm r})t + (\phi_0 + \phi_{\rm r})]}.
\end{equation}
The function generator also provides a phase shifted output $V_{\rm ref2}(t) = V_{\rm r} \cos{(2 \pi \nu_{\rm r} t + \phi_{\rm r})}$.
Thus, the four signals, namely $V_{\rm X}(t)$, $V_{\rm Y}(t)$, $V_{\rm ref1}(t)$ and $V_{\rm ref2}(t)$
are sent to a module which synthesizes a signal for feedback.
In a normal feedback mode, the feedback module executes arithmetic,
$V_{\rm X}(t) V_{\rm ref2}(t) - V_{\rm Y}(t) V_{\rm ref1}(t)$,
and provides a control voltage $V_{\rm FB}(t)$ according to which the feedback field $B_{\rm FB}(t)$ is generated:
\begin{align}
    V_{\rm FB} &= V_{\rm X}(t) V_{\rm ref2}(t) - V_{\rm Y}(t) V_{\rm ref1} \nonumber \\
               &= \frac{1}{2} V_0 V_{\rm r}^2 \cos{(2 \pi \nu_0 t + \phi_0)}.
    \label{eq:normal_FB}
\end{align}
Thus the phase of $V_{\rm FB}(t)$ has been shifted by 90$^\circ$ from $V_{\rm sig}(t)$
while the amplitude of $V_{\rm FB}$ is proportional to $V_{\rm sig}(t)$.
In a fixed-amplitude mode, $V_{\rm FB}(t)$ in Eq.(\ref{eq:normal_FB})
divided by square root of $V_{\rm X}^2 + V_{\rm Y}^2$
is adopted as a control signal for ${\bm B}_{\rm FB}(t)$:
\begin{align}
    V_{\rm FB}^{\rm (fixed)}(t) &= \frac{V_{\rm X}(t) V_{\rm ref2}(t) - V_{\rm Y}(t) V_{\rm ref1}}{\sqrt{V_{\rm X}^2 + V_{\rm Y}^2}} \nonumber \\
                    &= V_{\rm r} \cos{(2 \pi \nu_0 t + \phi_0)}.
    \label{eq:fixed_FB}
\end{align}
Thus the amplitude of $V_{\rm FB}^{\rm (fixed)}$ is made independent of the amplitude of the actual precession signal.

A feedback field according to Eq.(\ref{eq:normal_FB}) or (\ref{eq:fixed_FB}) is produced by a coil wound around the cell.
The transverse polarization $P_{\rm T}(t)$ decays due to relaxation effects,
while the torque generated by the feedback field tends to tilt the magnetization vector
from the $z$-axis to the $xy$-plane, thus enhancing $P_{\rm T}(t)$.
When two effects balance against each other,
the Xe spin precession is maintained far beyond the spin decoherence time,
avoiding the decay of $P_{\rm T}(t)$.

The $V_{\rm X}(t)$ and $V_{\rm Y}(t)$ signals from the second LAs
(four LA modules were employed, each for generation of the feedback fields for $^{129}$Xe and $^{131}$Xe,
and data taking for $^{129}$Xe and $^{131}$Xe) and
several parameters which represent the experimental condition
($i.e.$ the cell temperature $T_{\rm cell}$,
the pumping laser power $P_{\rm pump}$ and frequency $\nu_{\rm pump}$,
the probe laser power $P_{\rm probe}$ and frequency $\nu_{\rm probe}$)
were digitalized by the 16-ch ADC (DASmini-E500, Kei Technos Corp.)
and recorded in a PC at a sampling rate of 20Hz.
The maser frequency is derived from a train of the measured phases $\phi (t)$
via a relation $\phi(t) = \tan^{-1}{(V_{\rm Y}(t)/V_{\rm X}(t))}$:
By fitting
a theoretical function, which is assumed to be linear in time $t$,
to the measured $\phi (t)$,
the maser frequency is determined.

In order to investigate the long-term stability of masers,
frequency responses (\textit{i.e.} susceptibilities) to parameters of the experiment,
$B_0$, $T_{\rm cell}$, $P_{\rm pump}$, $\nu_{\rm pump}$, $P_{\rm pump}$ and $\nu_{\rm probe}$, were measured.
The frequency of the maser was measured by changing manually the above parameters one by one.
By combining the susceptibilities thus obtained and the measured instabilities of the individual parameters,
contributions from drifts in the individual parameters to the frequency instability are evaluated.

\section{Result and discussion}
\label{sec:exp}
\begin{figure}[htb]
    \centering
    \includegraphics[width=8.8cm]{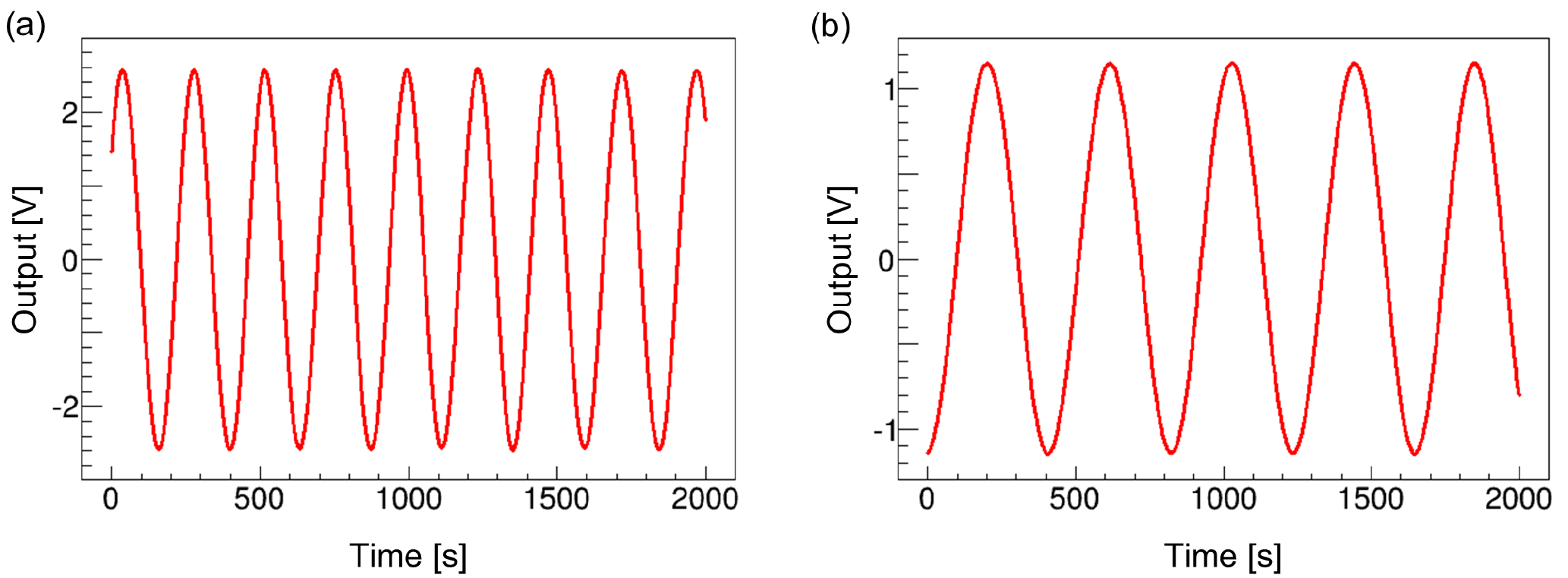}
    \caption{Spin oscillation signals for (a)$^{129}$Xe and (b)$^{131}$Xe.
            Beat signals from the lock-in amplifiers whose frequency was down converted to about few mHz
            were observed upon the reference frequency of 10.960~Hz for $^{129}$Xe and 3.253~Hz for $^{131}$Xe.}
    \label{fig:maser_oscilation}
\end{figure}

The observation of a $^{131}$Xe maser, and also its concurrent operation with a $^{129}$Xe maser,
have been performed in the present work for the first time.
The outputs from the lock-in amplifiers in the maser oscillations of the co-located $^{129}$Xe and $^{131}$Xe nuclear spins
are shown in Fig.~\ref{fig:maser_oscilation}.

\begin{figure}[htbp]
    \centering
    \includegraphics[width=8.8cm]{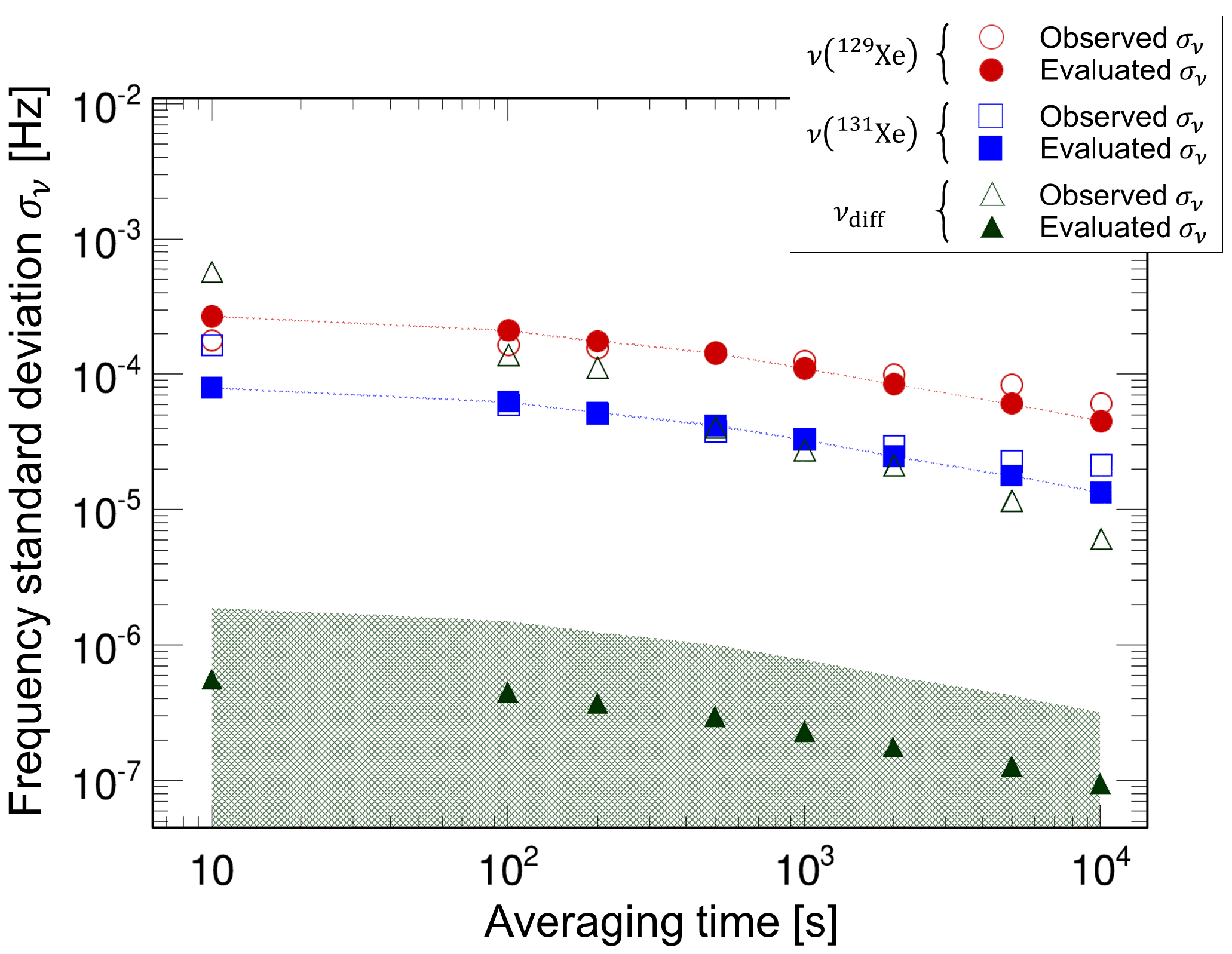}
    \caption{Long-term stability of masers and the evaluated contribution
            from the instabilities of the magnetic fields and the cell temperature.
            Red $\CIRCLE$, blue $\blacksquare$ and green $\blacktriangle$ symbols represent
            the evaluated contributions from the frequency instabilities due to magnetic effects
            for $\nu(^{129}{\rm Xe})$, $\nu(^{131}{\rm Xe})$ and $\nu_{\rm diff}$, respectively. 
            Hatched band represents the statistical error associated to the evaluated $\sigma_{\nu}$ for $\nu_{\rm diff}$.
            Note that the error bands corresponding to $\CIRCLE$ and $\blacksquare$ are so narrow
            that they appear as lines on the panel.
            Red $\Circle$, blue $\square$ and green $\triangle$ symbols represent
            the standard deviations of the measured $\nu(^{129}{\rm Xe})$, $\nu(^{131}{\rm Xe})$ and $\nu_{\rm diff}$, respectively.}
  \label{fig:maser_temp_dep}
\end{figure}

In order to assess the performance of the comagnetometry with the co-located $^{131}$Xe,
we define the differential frequency $\nu_{\rm diff}$ as
\begin{equation}
    \nu_{\rm diff} = \nu(^{129}{\rm Xe}) - \frac{\gamma(^{129}{\rm Xe})}{\gamma(^{131}{\rm Xe})}\nu(^{131}{\rm Xe}),
\end{equation}
where $\nu(^{129}{\rm Xe})$ and $\nu(^{131}{\rm Xe})$ are the maser frequencies for $^{129}$Xe and $^{131}$Xe,
and $\gamma(^{129}{\rm Xe})$/$\gamma(^{131}{\rm Xe}) = 3.37340(4)$~\cite{Brinkmann1963}
is the ratio between the gyromagnetic ratios for $^{129}$Xe and $^{131}$Xe.
Thus, $\nu_{\rm diff}$ represents the difference between the bare $^{129}$Xe frequency and $^{129}$Xe frequency
corrected for changes in the magnetic effects
(such as the field drift and the changes in the Rb density and polarization) 
which are inferred from the observed change in the co-located $^{131}$Xe spins. 
The evaluated contributions from drifts of the individual parameters
to the standard deviation of the maser frequencies determined in an averaging time of $10^4$~s
are listed in Table~\ref{tab:maser_params_dep}.
The evaluated standard deviation of the maser frequency due to the drifts in $T_{\rm cell}$,
$B_0$ and residual field $B_{\rm res}$,
which as a whole are expected to be eliminated by the $^{131}$Xe comagnetometer,
is plotted as a function of the averaging time in Fig.~\ref{fig:maser_temp_dep}.
\hdashlinegap=2pt 
\begin{table*}[htb]
  \centering
  \begin{tabular}{lr@{.}l@{\hspace{0pt}}c@{\hspace{0pt}}lr@{.}l@{\hspace{0pt}}c@{\hspace{0pt}}lr@{.}l@{\hspace{0pt}}c@{\hspace{0pt}}l} \hline \hline
    Parameter &\multicolumn{4}{c}{$\sigma_\nu(^{129}{\rm Xe})$ $[\mu{\rm Hz}]$} & \multicolumn{4}{c}{$\sigma_\nu(^{131}{\rm Xe})$ $[\mu{\rm Hz}]$} & \multicolumn{4}{c}{$\sigma_\nu(\rm diff)$ $[\mu{\rm Hz}]$} \\ \hline
    Static field $B_0$                          & 0&0145&$\pm$&0.0001 & 0&00427&$\pm$&0.0005    & 0&000205&$\pm$&0.000131 \\
    Residual field $B_{\rm res}$                & 0&454&$\pm$&0.004   & 0&133&$\pm$&0.001       & 0&00640&$\pm$&0.00410   \\
    Cell temperature $T_{\rm cell}$             & 45&4&$\pm$&0.6      & 13&4&$\pm$&0.2          & 0&095&$\pm$&0.224       \\
    Pumping laser power $P_{\rm pump}$          & 8&27&$\pm$&0.29     & 1&96&$\pm$&0.10         & 1&64&$\pm$&0.15         \\
    Pumping laser frequency $\nu_{\rm pump}$    & 25&4&$\pm$&0.6      & 7&73&$\pm$&0.13         & $<$0&150                \\
    Probe laser power $P_{\rm probe}$           & 0&241&$\pm$&0.041   & 0&127&$\pm$&0.022       & 0&186&$\pm$&0.063       \\
    Probe laser frequency $\nu_{\rm probe}$     & 0&0883&$\pm$&0.0177 & 0&0265&$\pm$&0.0583     & $<$0&0159               \\ \hdashline
    Quadrature sum          & 52&7&$\pm$&0.9      & 15&6&$\pm$&0.3          & 1&66&$\pm$&0.32         \\ \hline
    Measured frequency      & \multicolumn{4}{c}{61.4} & \multicolumn{4}{c}{21.6} & \multicolumn{4}{c}{6.15} \\ \hline \hline
  \end{tabular}
  \caption{Evaluated contributions from instabilities of various parameters in the experiment
            to the standard deviations for $\nu(^{129}{\rm Xe})$, $\nu(^{131}{\rm Xe})$ and $\nu({\rm diff})$
            for an averaging time $10^4$~s.}
  \label{tab:maser_params_dep}
\end{table*}

Figure~\ref{fig:maser_temp_dep} reveals that, for both $^{129}$Xe and $^{131}$Xe cases,
the standard deviation $\sigma_{\nu}$ for the measured maser frequency
is mostly explained by the evaluated contributions
from drifts in $B_0$, $B_{\rm res}$ and $T_{\rm cell}$.
Since $B_0$, $B_{\rm res}$ and $T_{\rm cell}$
(note that change in $T_{\rm cell}$ causes change in the Rb number density in the cell,
hence change in the effective magnetic field acting on the Xe spin via the Rb-Xe nucleus Fermi contact interaction)
both alter the maser frequency by a factor common to the two species $^{129}$Xe and $^{131}$Xe,
their effects should be compensated in $\nu_{\rm diff}$.
In fact, the evaluated $\sigma_{\nu}$ for $\nu_{\rm diff}$
has been dramatically reduced from those for $^{129}$Xe and $^{131}$Xe frequencies themselves
(from $\sim$45~$\mu$Hz to below $\sim$320~nHz at the averaging time of $10^4$~s),
thus demonstrating clearly that the proposed comagnetometry
using $^{131}$Xe co-located with $^{129}$Xe works quite efficient.

In Table~\ref{tab:maser_params_dep},
individual contributions to the standard deviations
(for an averaging time of $10^4$~s) of the maser frequencies $\nu(^{129}{\rm Xe})$, $\nu(^{131}{\rm Xe})$ and $\nu_{\rm diff}$
that arise from drifts in the experimental parameters are evaluated
from the measured susceptibilities of the frequencies to the respective parameters.
One may notice from the Table that, although the contributions
from drifts in $B_0$, $B_{\rm res}$ and $T_{\rm cell}$ are largely reduced in $\sigma_\nu({\rm diff})$ 
as compared to those in $\sigma_\nu(^{129}{\rm Xe})$ and $\sigma_\nu(^{129}{\rm Xe})$ as stated earlier,
contributions of other drifts such as $P_{\rm pump}$, $P_{\rm probe}$, $\nu_{\rm pump}$ and $\nu_{\rm probe}$ drifts
seem to remain essentially unaltered.
This gave us a hint on an additional, and previously unrecognized, source of frequency drift
which makes the evaluated $\nu_{\rm diff}$ (1.66 $\mu$Hz) substantially larger
than that evaluated for the magnetic contributions only.
Indeed, one may notice that the interaction of ${\bm B}_{\rm FB}(t)$ with the Rb magnetization
(which serves as an indicator of the Xe spin direction in the optical detection of spin)
does not enter explicitly the equations of motion, Eqs.(\ref{eq:Bloch_x})-(\ref{eq:Bloch_z}),
although it seems to actually influence the time evolution of ${\bm P}(t)$.
A quantitative analysis of this effect, its experimental verification
and a scheme under our current development by which this effect will be removed
will be presented in a forthcoming publication.

Finally, the \textit{observed} value of $\sigma_\nu ({\rm diff})$ in a $10^4$~s averaging time
(the rightmost green $\triangle$ symbol in Fig.~\ref{fig:maser_temp_dep}) is 6.15~$\mu$Hz.
The frequency instability of the spin maser
with an external feedback scheme is determined
from the phase error in the optical detection of spin precession:
The optical detection of spin precession involves a phase error $\sigma_\phi$,
which is governed by the SNR of the PD output signal, as
\begin{equation}
    \sigma_\phi = \frac{1}{SNR} \sqrt{\frac{\Delta t}{T_{\rm m}}},
    \label{eq:phase_dev}
\end{equation}
where $T_{\rm m}$ is the averaging time.
The statistical fluctuation (of size $\sigma_\phi$) around zero of the angle
${\bm P}_{\rm T}(t)$ and ${\bm B}_{\rm FB}(t)$
then introduces a fluctuation in the spin precession frequency
via the pulling effect, resulting in a standard deviation of the maser frequency
\begin{equation}
    \sigma_\nu = \frac{\tan{\sigma_\phi}}{2 \pi T_2}.
\end{equation}
In fact, inserting the current experimental conditions,
$SNR =$ 60, $\Delta t =$ 0.05~s, $T_{\rm m} = T_2 =$ 20~s
and $\sigma_\phi =$ 0.00083~rad
leads to $\sigma_{\nu}$ of 6.6~$\mu$Hz for a $10^4$~s averaging time,
providing a reasonable explanation for the observed $\sigma_\nu$ of 6.2~$\mu$Hz.
The shot-noise-limit of SNR for the current experimental setup is calculated to be about 1100.
By narrowing the measurement bandwidth from 50~kHz
(which is currently determined by the resonance frequency of the commercial PEM) to 70~Hz,
the shot-noise-limited SNR can be increased to about 46000.
This value of SNR corresponds to the standard deviation of $\approx$ 1~nHz
for $\nu_{\rm diff}$ for an averaging time of $10^6$~s ($\sim$12~days).

\section{Summary}
\label{sec:summary}
A nuclear spin maser of $^{131}$Xe has been successfully operated in concurrence
with a $^{129}$Xe spin maser, the two masers being co-located in a common cell.
This accomplishment implies the possibility of application of a $^{131}$Xe spin maser
as an efficient comagnetometer for $^{129}$Xe precession measurements,
such as experiments searching for atomic EDMs or for hypothetical oscillatory or transient CP-odd background fields in space.
Indeed, the present analysis indicates that frequency drifts
due to ambient magnetic effects on $^{129}$Xe spin are reduced by two orders of magnitude
by applying the appropriate correction based on the measured change in the $^{131}$Xe spin precession frequency.
The present work, however, has also unveiled the presence of another source of maser frequency instability
that stems from drifts in the pumping and probe light powers.
Experimental verification of the mechanism inferred for this phenomenon,
implementation of a scheme to remove the frequency drift caused by it,
and improvement of the SNR for the precession phase detection are the subjects of our current development,
aiming at the realization of nHz level precision in a $10^6$~s long measurement for the precession frequency.

\section*{Acknowledgements}
This work was supported by the JSPS KAKENHI (Grant-in-Aid for Scientific Research) 21104004, 21244029, 26247036, 17K14310;
RIKEN Incentive Research Grant; and RIKEN SPDR research funding.


\begin{thebibliography}{99}
\bibitem{Tullney2013}
K. Tullney {\em et al.}, Physical Review Letters {\bf 111} (2013) 100801, arXiv:1303.6612.

\bibitem{Bulatwicz2013}
M. Bulatwicz2013 {\em et al.}, Physical Review Letters {\bf 111} (2013) 102001, arXiv:1301.5224.

\bibitem{Vasilakis2009}
G. Vasilakis {\em et al.}, Physical Review Letters {\bf 103} (2009) 261801, arXiv:0809.4700.

\bibitem{Bear2000}
D. Bear {\em et al.}, Physical Review Letters {\bf 85} (2000) 5038--5041.

\bibitem{Allmendinger2014}
F. Allmendinger {\em et al.}, Physical Review Letters {\bf 112} (2014) 110801, arXiv:1312.3225.

\bibitem{Rosenberry2001}
M. A. Rosenberry, T. E. Chupp, Physical Review Letters {\bf 86} (2001) 22-25.

\bibitem{Graner2016}
B. Graner {\em et al.}, Physical Review Letters {\bf 116} (2016) 161601, arXiv:1601.04339.

\bibitem{Pendlebury2015}
J. M. Pendlebury {\em et al.}, Physical Review D {\bf 92} (2015) 092003, arXiv:1509.04411.

\bibitem{Walker1997}
T. G. Walker and W. Happer, Reviews of Modern Physics {\bf 69} (1997) 629--642.

\bibitem{Yoshimi2002}
A. Yoshimi {\em et al.}, Physics Letters A {\bf 304} (2002) 13--20.

\bibitem{Yoshimi2012}
A. Yoshimi {\em et al.}, Physics Letters A {\bf 376} (2012) 1924--1929.

\bibitem{Ichikawa2014}
Y. Ichikawa {\em et al.}, EPJ Web of Conferences {\bf 66} (2014) 05007.

\bibitem{Sato2015}
T. Sato {\em et al.}, JPS Conference Proceedings {\bf 6} (2015) 020031.

\bibitem{Inoue2016}
T. Inouse {\em et al.}, European Physical Journal D {\bf 70} (2016) 129.

\bibitem{Flambaum2016}
V.V. Flambaum, Physical Review Letters {\bf 117} (2016) 072501, arXiv:1603.05753.

\bibitem{Budker2014}
D. Budker {\em et al.}, Physical Review X {\bf 4} (2014) 021030, arXiv:1306.6089.

\bibitem{Abel2017}
C. Abel {\em et al.}, Physical Review X {\bf 7} (2017) 041034, arXiv:1708.06367.

\bibitem{Heil2013}
W. Heil {\em et al.}, Annalen der Physik {\bf 525} (2013) 539--549.

\bibitem{Schaefer1989}
S. R, Schaefer {\em et al.}, Physical Review A {\bf 39} (1989) 5613.

\bibitem{Ma2011}
Z. L. Ma, E. G. Sorte, B. Saam, Physical Review Letters {\bf 106} (2011) 193005.

\bibitem{Romalis1998}
M. V. Romalis, G. D. Cates, Physical Review A {\bf 58} (1998) 3004--3011.

\bibitem{Kwon1981}
T. M. Kwon, J. G. Mark, C. H. Volk, Physical Review A {\bf 24} (1981) 1894--1903.

\bibitem{Stupic2011}
K. F. Stupic {\em et al.}, Journal of Magnetic Resonance {\bf 208} (2011) 58--69.

\bibitem{Sakamoto2015}
Y. Sakamoto {\em et al.}, Hyperfine Interactions {\bf 230} (2015) 141-146.

\bibitem{Green1998}
K. Green {\em et al.}, Nuclear Instruments and Methods A {\bf 404} (1998) 381-393.

\bibitem{Brinkmann1963}
D. Brinkmann, Helvetica Physica Acta {\bf 36} (1963) 413--414.

\end{thebibliography}
\end{document}